# A Study of Rarely Appeared Instructions in an Executable Binary

무로도바 노지마[*], 구형준[*]

성균관대학교 소프트웨어학과 (대학원생[*], 교수[**])

Nozima Murodova[*], Hyungjoon Koo[**]

*Department of Computer Science and Engineering, Sungkyunkwan University

## Abstract

An executable binary typically contains a large number of machine instructions. Although the statistics of popular instructions is well known, the distribution of non-popular instructions has been relatively under explored. Our finding shows that an arbitrary group of binaries comes with both i) a similar distribution of common machine instructions, and ii) quite a few rarely appeared instructions (e.g., less than five occurrences) apart from the distribution. Their infrequency may represent the signature of a code chunk or the footprint of a binary. In this work, we investigate such rare instructions with an in-depth analysis at the source level, classifying them into four categories.

## I. INTRODUCTION

An executable binary is the final product of a compilation toolchain that entails complex transformations from a source code. The binary contains a series of machine instructions along with data generated by a compiler. The instructions are chosen during an instruction selection phase as part of compilation. It is well known that not every machine instruction accounts for a similar proportion although there are a large number of instructions [8]. In particular, a complex instruction set computer (CISC) like x86 provides a single instruction that is capable of executing multiple-step operations. This inevitably brings about an imbalanced occurrence of each instruction.

Unlike popular instruction statistics [8], the distribution of non-popular instructions has been relatively under explored. However, relying on the frequency of an instruction merely by counting is problematic because the operand(s) of the instruction may vary (e.g., 4-byte immediate can hold 2^32 values), rendering occurrence-based statistics meaningless.

Recent efforts to be able to better infer the contextual semantics from a binary often take advantage of varying techniques in the field of natural language processing (NLP). While NLP generates an embedding per each word (e.g., word2vec) before feeding it to build a model, care must be taken when applying it to a raw instruction due to possibly too many tokens for training as well as out-of-vocabulary (OOV) for testing. To mitigate this problem, an instruction is often normalized as a pre-processing step before being vectorized. This helps to reduce the number of overall vocabularies (tokens).

DeepSemantic [2] proposes a well-balanced normalization technique to strike a balance by maintaining a reasonable number of tokens for building a model while preserving the original semantics of an instruction. According to the normalization with their dataset, the whole instructions can be represented with 17,225 vocabularies (tokens). Further, 144 tokens account for around 84% of the whole (normalized) instructions. Our interest is rarely seen instructions (approximately 9k) that form a long tail.

In this work, we first confirm that the distribution of such infrequently appeared instructions (i.e., five or less) is quite consistent in an arbitrary binary set (Figure 1), and then investigate them with source codes. Our major finding shows that it is possible that a certain instruction can represent the footprint of a code chunk (e.g., function, binary). We classify rarely-seen instructions into the following four categories: i) an instruction that specifies a compiler, ii) an instruction for accessing a member variable of a structure, iii) an instruction to support a certain floating-point operation, and iv) an instruction from a manually written assembly.

## II. RARELY APPEARED INSTRUCTIONS

**Experimental Setup and Dataset.** With SPEC2017 benchmark test suites [3] under a x86_64 environment, we compiled them with two compilers (e.g., gcc, clang), four optimization levels (e.g., O0-O3), and a debugging flag enabled. In total, our dataset consists of 120 binaries, which contains 754,833 functions, and 11,929 unique normalized instructions. Table 1 describes the distribution of rare instructions.

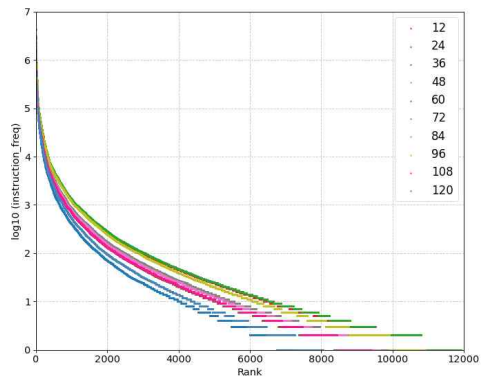

**Figure 1.** A log scale of an instruction frequency (y-axis) for a subset of our corpus based on a rank by the most common ones (x-axis), indicating a consistant ratio of rare instructions.

Note that we use normalized instructions. In order to confirm the common distribution of these rare instructions for different sets of binary data, we splitted the whole corpus into 10 subsets. Figure 1 shows the relationship of instruction frequencies and their ranking for each data subsets (12-120), creating a long tail for rare instructions. It indicates that we can consistently observe such instructions in any arbitrary dataset.

| Instruction freq. | Number | Ratio (%) |
|---|---|---|
| 1 | 1133 | 0.09 |
| 2 | 1276 | 0.10 |
| 3 | 707 | 0.06 |
| 4 | 580 | 0.05 |

**Table 1.** Number of rarely appeared instructions (<5) from the whole corpus.

**Instruction-source Mapping.** For further analysis, we automatically extract the address information during instruction normalization and utilize them with addr2line utility in order to inspect the matching line in the particular source code.

**Experimental Analysis.** As our corpus is compiled with different optimization levels, in order to inspect the change on the assembly instruction we manually analyze the target instruction for each binary optimization level. In some cases the instructions are not found when compiled with different optimization level which makes them appear rarely. According to our observations, they sometimes are connected with memory access [3] instruction to a particular custom data type (structure) which can serve as a unique identifier or signature for that piece of the program. This phenomenon indicates high possibility of rare instructions to represent a special birthmark of a binary.

**Category.** We study rare instructions that have been appeared less than five times in our dataset as follow: a) a compiler-specific instruction (i.e compiler intrinsics) [6], b) an instruction for accessing a member variable in a particular structure, c) an instruction to support a floating-point operation, and d) an instruction from a manually written assembly for a special purpose.

## III. RESULTS

**Case study 1.** Instructions that belong to this group mainly contain Intel C/C++ compiler intrinsics [5] which is specific for each compiler. For example, instruction *kortestw_k1_k0* seen 3 times belongs to gcc only, which performs a bitwise OR between the vector mask register k0, and the vector mask register k1 [7].

According to Intel reference [1], kortestw originally has the equivalent to compiler intrinsic function *KORTESTW__mmask16_mm512_kortest[cz](__mask16 a, __mmask 16 b)*. This instruction was used in 2 different functions of the same binary with optimization level 02 and 03 only. In the first function *quant_2x2_dc* the estimated matching block is shown in Listing 1.

```
1. #define QUANT_ONE( coef, mf, f )
2. {
3.  if((coef) > 0)
4.      (coef) = (f + (coef)) * (mf) >> 16;
5.  else
6.  (coef) = - ((f - (coef)) * (mf) >> 16);
7.  nz |= (coef);
8. }
```

**Listing 1. kortestw_k1_k0 assembly instruction matching code block - line 7 (first occurrence)**

Listing 2 shows an example of the instruction *zigzag_interleave_8x8_cavlc* implementing logical OR operation in the line number 5.

```
1. static void
   zigzag_interleave_8x8_cavlc(int16_t
   *dst,int16_t *src,uint8_t *nnz ){
2. for(int i =0; i <4; i++){
3.     int nz =0;
4.     for(int j =0; j <16; j++){
5.         nz |= src[i+j*4];
6.         dst[i*16+j]= src[i+j*4];
7.     }
8.     nnz[(i&1)+(i>>1)*8]=!!nz;
9. }
```

**Listing 2. kortestw_k1_k0 assembly instruction matching code block (second occurrence)**

Other similar examples can be seen for instructions like *adc_bp8_qwordptr[sp8+disp], korb_k2_k0_k1, blsi_bp4_reg4*, etc.

**Case study 2.** We observed and confirmed that most instructions pointing to particular structure member variables were also seen only a few times. Some of these examples include v-instructions *(vpmovzxbq, vbroadcastss* and many more) which are only used with large size registers such as xmm, ymm. These opcodes mainly belong to the group of EVEX encoded instructions by Intel which also come with a set of compiler equivalent intrinsics. Once they are met in gcc compiled binary, it is very unlikely they can be seen in the clang output as well. Now let us look into that scenario closer.

Some of the interesting instructions can be found in gcc binaries (which are not found with clang), appending opcode vbroadcastss: (1) *vbroadcastss_regymm_regxmm* and (2) *vbroadcastss_regxmm_dwordptr[sp8+8]*.

Listing 3 shows the code snippet with the above rare with the combination of other common instructions pointing to the line 1 (which is the function start)

```
1. void normalize_dq(DualQuat *dq,float
        totweight){
2.      const float scale =1.0f /
        totweight;
```

**Listing 3.** *vbroadcastss_regymm_regxmm* **assembly instruction matching code block**

where the corresponding struct type is given in Listing 4.

```
1. typedef struct DualQuat {
2.      float quat[4];
3.      float trans[4];
4.      float scale[4][4];
5.      float scale_weight;
6. } DualQuat;
```

**Listing 4.** *DualQuat* **struct definition**

Listing 5 and 6 show one more example similar to the one above with the same operation code vbroadcastss, and the definition of the *shadfac*.

```
1. if (isec->mode==RE_RAY_SHADOW_TRA){
2.      shadfac[0]/=div;
3.      shadfac[1]/=div;
4.      shadfac[2]/=div;
5.      shadfac[3]/=div;
6. }
```

**Listing 5.** *vbroadcastss_regxmm_dwordptr[sp8+8]* **assembly instruction matching code block (line 2)**

```
1. typedef struct LampShadowSubSample {
2.      int samplenr;
3.      float shadfac[4];
4. } LampShadowSubSample;
```

**Listing 6.** *LampShadowSubSample* **struct definition (line 3)**

We observe that this applies to compiler specific instructions as well but refer to a particular data type. So far we made sure that these instructions belong to specific compiler's binary only (either gcc or clang). This is the way how these compilers interpret certain instructions differently.

**Case study 3.** Our finding shows that new instructions for supporting floating point operations which are rarely used by compilers and represent SIMD (Single Instruction Multiple Data) instruction sets. Examples are found with instructions such as *vcmplesd_regxmm_regxmm_qwordptr[sp8+8]* for Comparing Scalar Double-Precision Floating-Point Value, *vcmpgtss_regxmm_regxmm_regxmm* for Comparing Scalar Single-Precision Floating-Point Value, *vcmpeq_usss_regxmm_regxmm_dwordptr[ip8+disp]* for Comparing Scalar Single-Precision Floating-Point Value. Compilers can use these EVEX encoded instructions for three operand instructions. EVEX (enhanced vector extension) schema is essentially a 4-byte extension to VEX schema which supports the AVX512 instruction set and allows addressing new 512 bit ZMM and new operand mask registers [7]. These instructions also come with their equivalent compiler intrinsics extending the Case 1 studies.

**Case study 4.** Although we have not observed many of these examples, which can be also addressed as programmer specific instructions, we found them to be much more interesting, as these instructions are written manually in assembly in order to avoid compiler auto generated intrinsics. One of these examples is *prefetcht0*, this instruction is used for caching, namely fetches the line of data from memory that contains the byte specified with source operand to a location in the cache hierarchy specified by locality hint T0 (temporal data) – prefetching data into all levels of cache hierarchy. Listing 7 shows the code line where developers implemented this instruction manually in order to avoid gcc built_in function to make it during compilation.

```
1. static ALWAYS_INLINE void
2. x264_prefetch(void *p ){
3.     asm volatile("prefetcht0
4.              %0"::"m"(*(uint8_t*)p));
5. }
```
**Listing 7.** Manually implemented *prefetch0* instruction

## V. LIMITATIONS & FUTURE WORK

So far limited analysis been held due to various case study results and complex dataset. Categories are selected with mostly occurred findings, but we believe there are other interesting cases which can be discovered with an automated analysis. Also, mapping the exact line number in source code should be checked for correctness (i.e, addr2line utility sometimes found to be unreliable). Making a fully automated tool for extracting rare instructions in source file according to instruction addresses would be a good future work for studying the importance of rare instructions.

## VI. CONCLUSION

We conduct an in-depth instruction level analysis on x86-64 binaries by taking a close look at rarely appeared instructions. Our finding shows that such instructions imply that a special operation, compiler intrinsic, structure or written assembly, which could be the birthmark of a binary. Further investigation is required to hypothesize our observation., which is part of our future work.

## [REFERENCES]